\documentclass[preprint,aps,showpacs,nofootinbib,preprintnumbers,amsmath,amssymb]{revtex4-1}
\usepackage{epsfig}

\usepackage{dcolumn}
\usepackage{bm}
\usepackage[usenames ,dvipsnames]{xcolor}
\usepackage{slashed}
\usepackage{graphicx,color}

\begin{document}
\title{Anti-triplet charmed baryon decays with SU(3) Flavor Symmetry}

\author{C.Q. Geng$^{1,2}$, Y.K. Hsiao$^{1,2}$, Chia-Wei Liu$^{2}$ and Tien-Hsueh Tsai$^{2}$}
\affiliation{
$^{1}$School of Physics and Information Engineering, Shanxi Normal University, Linfen 041004, China\\
$^{2}$Department of Physics, National Tsing Hua University, Hsinchu, Taiwan 300
}\date{\today}

\begin{abstract}
We study the decays of the
anti-triplet charmed baryon state  $(\Xi_c^0,\Xi_c^+,\Lambda_c^+)$
based on the $SU(3)$ flavor symmetry.
In particular, after predicting
 ${\cal B}(\Xi_c^0\to \Xi^-\pi^+)=(15.7 \pm 0.7)\times 10^{-3}$
and ${\cal B}(\Xi_c^+\to\Xi^-\pi^+\pi^+)=(14.7\pm 8.4)\times 10^{-3}$,
we  extract that
${\cal B}(\Xi_c^0\to\Lambda K^-\pi^+,\Lambda K^+K^-,\Xi^- e^+\nu_e)=(16.8\pm 2.3,0.45\pm 0.11,48.7\pm 17.4)\times 10^{-3}$
and
${\cal B}(\Xi_c^+\to p K_s^0 K_s^0,\Sigma^+ K^-\pi^+,\Xi^0\pi^+\pi^0,\Xi^0 e^+\nu_e)=
(1.3\pm 0.8,13.8\pm 8.0,33.8\pm 21.9,33.8^{+21.9}_{-22.6})\times 10^{-3}$.
We also find that ${\cal B}(\Xi_c^0\to \Xi^{0} \eta,\Xi^{0} \eta')=
(1.7 ^{+ 1.0 }_{- 1.7 },8.6 ^{+ 11.0 }_{-\;\,6.3 })\times 10^{-3}$,
${\cal B}(\Xi_c^0\to \Lambda^{0} \eta,\Lambda^{0} \eta')
=(1.6 ^{+ 1.2 }_{- 0.8 },9.4 ^{+ 11.6 }_{-\;\,6.8 })\times 10^{-4}$ and
${\cal B}(\Xi_c^+\to \Sigma^{+} \eta,\Sigma^{+} \eta')
=(28.4 ^{+ 8.2 }_{- 6.9 },13.2 ^{+ 24.0 }_{- 11.9 })\times 10^{-4}$.
 These $\Xi_c$ decays with  the branching ratios of  $O(10^{-4}-10^{-3})$
 are clearly promising to be observed by the BESIII and LHCb experiments.

\end{abstract}

\maketitle
\section{Introduction}
In terms of the $SU(3)$ flavor ($SU(3)_f$) symmetry,
the $\Xi_c$ decays should be in association with the $\Lambda_c^+$ ones
as $\Xi_c^0$, $\Xi_c^+$ and $\Lambda_c^+$ are united as
the lowest-lying anti-triplet of the charmed baryon states (${\bf B}_c$).
%
Nonetheless, in accordance with $f_{\Xi_c^+}+f_{\Xi_c^0}+f_{\Omega_c^0}
\simeq 0.136\,f_{\Lambda_c^+}$ estimated in Refs.~\cite{Lisovyi:2015uqa,Alexander:1996wy},
where $f_{{\bf B}_c,\Omega_c^0}$ 
stand for the fragmentation fractions for the rates of the charmed baryon productions,
the measurements of the $\Xi_c$ decays are not  easy tasks
compared to the $\Lambda_c^+$ ones.
For example,
the two-body $\Lambda_c^+\to {\bf B}_n M$ decays
with ${\bf B}_n(M)$ the baryon (pseudoscalar-meson ) have been extensively studied 
by experiments.
Interestingly,  six decay $\Lambda_c^+$ decay modes have been recently reexamined or measured
by BESIII~\cite{Ablikim:2015flg,Ablikim:2017ors}.
%
In addition,
LHCb has just observed
the three-body $\Lambda_c^+\to p MM$ decays~\cite{Aaij:2017rin},
together with their CP violating asymmetries~\cite{Aaij:2017xva}.
%
 However,  no much progress has been made in the $\Xi_c$ decays.
In particular, none of the absolute branching fractions in the $\Xi_c$ decays
has been given yet.
Instead,  these   decays are experimentally measured by
relating  the decays of $\Xi_c^+\to \Xi^-\pi^+\pi^+$ or $\Xi_c^0\to \Xi^-\pi^+$, 
and can only be determined once $f_{\Xi_c^{0,+}}$~\cite{pdg} are known.

Since BESIII and LHCb are expected to search for all possible anti-triplet charmed baryon decays,
one can test whether or not the studies of $\Lambda_c^+\to {\bf B}_n M$
can be applied to $\Xi_c^{0,+}\to {\bf B}_n M$.
Theoretically, the factorization  for
the $b$~baryon decays~\cite{Hsiao:2014mua,Hsiao:2017tif,Hsiao:2016hrk,
Geng:2016gul,Geng:2016drz,Hsiao:2015txa}
does not work for the charmed baryon decays,
which receive corrections by taking into account
the nonfactorizable effects~\cite{Cheng:1991sn,Cheng:1993gf,
Zenczykowski:1993hw,Fayyazuddin:1996iy,Dhir:2015tja,Cheng:2018hwl}.
On the other hand, the possible
$b$ or $c$ hadron decay modes can be examined by
the $SU(3)_f$ symmetry~\cite{Lu:2016ogy,
He:2000ys,Fu:2003fy,Hsiao:2015iiu,He:2015fwa,He:2015fsa,
Savage:1989qr,Savage:1991wu,h_term,Wang:2017azm,
Geng:2017esc,Geng:2017mxn}.
Furthermore,
 the symmetry approach has been extended
to explore the doubly and triply charmed baryon decays~\cite{Geng:2017mxn},
which helps to establish the spectroscopies of
the doubly and triply charmed baryon states~\cite{Aaij:2017ueg},
such as the to-be-confirmed $\Xi_{cc}^+$ state~\cite{Mattson:2002vu,Ocherashvili:2004hi,
Ratti:2003ez,Aubert:2006qw,Chistov:2006zj,Aaij:2013voa}.

Moreover, to test the validity of the $SU(3)_f$ symmetry in the anti-triplet charmed baryon decays,
a complete numerical analysis for the decays is necessary.
In fact, the decays of $\Lambda_c^+\to {\bf B}_n M$ have been explained well 
by the global fit in Ref.~\cite{Geng:2017esc},
together with the predictions of
${\cal B}(\Xi_c^+\to\Xi^0\pi^+)=(8.0\pm 4.1)\times 10^{-3}$ and
${\cal B}(\Xi_c^0\to\Lambda^0\bar K^0)=(8.3\pm 0.9)\times 10^{-3}$,
in agreement with the values of $(7.2\pm 3.5,8.3\pm 3.7)\times 10^{-3}$
extracted from the ratios of
${{\cal B}(\Xi_c^+\to\Xi^0\pi^+)}/{{\cal B}(\Xi_c^+\to\Xi^0 e^+\nu_e)}$ and
${{\cal B}(\Xi_c^0\to\Lambda^0\bar K^0)}/{{\cal B}(\Xi_c^0\to\Xi^- e^+\nu_e)}$,
respectively~\cite{Geng:2017mxn}.
%

In this report,
we will systematically study
the two-body weak $\Xi_c\to {\bf B}_n M$ decays based on the $SU(3)_f$ symmetry and give some specific numerical results,
which can be tested in the future measurements by BESIII and LHCb.
By taking the predicted ${\cal B}(\Xi_c^0\to \Xi^-\pi^+)$ as the theoretical input,
we will also estimate the branching ratios of  other $\Xi_c$ decays in the PDG~\cite{pdg}, which
 are related  to $\Xi_c^0\to \Xi^-\pi^+$.

\section{Formalism}
For the two-body anti-triplet of the lowest-lying charmed baryon decays of ${\bf B}_c\to {\bf B}_n M$,
where ${\bf B}_{c}=(\Xi_c^0,-\Xi_c^+,\Lambda_c^+)$
and ${\bf B}_n$ ($M$) are the baryon (pseudoscalar) octet states,
the effective Hamiltonian responsible for
the tree-level $c\to  s u\bar d$, $c\to u q\bar q$ and $c\to  du\bar s$ transitions
are given by~\cite{Buras:1998raa}
\begin{eqnarray}\label{Heff}
{\cal H}_{eff}&=&\sum_{i=+,-}\frac{G_F}{\sqrt 2}c_i
\left(V_{cs}V_{ud}O_i+V_{cd}V_{ud} O_i^\dagger+V_{cd}V_{us}O'_i\right),
\end{eqnarray}
with $q\bar q=d\bar d$ or $s\bar s$,
$G_F$  the Fermi constant, $V_{ij}$ the CKM matrix elements, and
$c_{\pm}$ the scale-dependent Wilson coefficients
to take into account the sub-leading-order QCD corrections.
The four-quark operators
$O_\pm^{(\prime)}$ and $O_\pm^\dagger\equiv O_\pm^d-O_\pm^s$
in Eq.~(\ref{Heff}) can be written as
\begin{eqnarray}\label{O12}
&&
O_\pm={1\over 2}\left[(\bar u d)_{V-A}(\bar s c)_{V-A}\pm (\bar s d)_{V-A}(\bar u c)_{V-A}\right]\,,\;\nonumber\\
&&
O_\pm^q={1\over 2}\left[(\bar u q)_{V-A}(\bar q c)_{V-A}\pm (\bar q q)_{V-A}(\bar u c)_{V-A}\right]\,,\;\nonumber\\
&&
O'_\pm={1\over 2}\left[(\bar u s)_{V-A}(\bar d c)_{V-A}\pm (\bar d s)_{V-A}(\bar u c)_{V-A}\right]\,,
\end{eqnarray}
where $(\bar q_1 q_2)_{V-A}=\bar q_1\gamma_\mu(1-\gamma_5)q_2$.
By using $(V_{cs}V_{ud},V_{cd}V_{ud},V_{cd}V_{us})\simeq (1,-s_c,-s_c^2)$
in Eq.~(\ref{Heff}) with $s_c\equiv \sin\theta_c=0.2248$~\cite{pdg}
representing  the well-known Cabbibo angle $\theta_c$, the decays with
$O_\pm$, $O_\pm^\dagger$ and$O_\pm^{\prime}$ are
the so-called Cabibbo-allowed, Cabibbo-suppressed, and
doubly Cabibbo-suppressed processes, respectively.
For instance of the Cabibbo-allowed decay,
${\cal B}(\Lambda_c^+\to p \bar K^0)=(3.16\pm 0.16)\times 10^{-2}$
is measured to be 50 times larger than
${\cal B}(\Lambda_c^+\to \Lambda K^+)=(6.1\pm 1.2)\times 10^{-4}$,
which is the Cabibbo-suppressed case,
whereas none of the doubly Cabibbo-suppressed ones has been observed~\cite{pdg}.

Without explicitly showing the Lorentz indices, 
the operators in Eq.~(\ref{O12}) behave as $(\bar q^i q_k \bar q^j)c$, 
with $q_i=(u,d,s)$ as the triplet of $3$, 
which can be decomposed as the irreducible forms 
under the $SU(3)_f$ symmetry, that is,
$(\bar 3\times 3\times \bar 3)c=(\bar 3+\bar 3'+6+\overline{15})c$.
Accordingly, $(O_-,O_+)$ fall into the irreducible presentations
of $({\cal O}_{6},{\cal O}_{\overline{15}})$, 
given by~\cite{Savage:1989qr}
\begin{eqnarray}
{\cal O}_{6}=\frac{1}{2}(\bar u d\bar s-\bar s d\bar u)c\,,
{\cal O}_{\overline{15}}=\frac{1}{2}(\bar u d\bar s+\bar s d\bar u)c\,,
\end{eqnarray}
which correspond to the tensor notations of
$1/2\epsilon^{ijl}H(6)_{lk}$ and $H(\overline{15})^{ij}_k$, respectively, 
with $(i,j,k)$  representing the quark indices and the non-zero entries being
$H_{22}(6)=2$ and $H_2^{13}(\overline{15})=H_2^{31}(\overline{15})=1$.
Note that $O_\pm^{\dagger}$ and $O_\pm^{\prime}$
also have similar irreducible representations,
resulting in the non-zero entries of 
$H_{23,32}(6)=-2s_c$, $H_2^{12,21}(\overline{15})=-H_3^{13,31}(\overline{15})=s_c$,
$H_{33}(6)=2s_c^2$, and $H_3^{12,21}(\overline{15})=-s_c^2$~\cite{Savage:1989qr}.
%
By using the bases of  the $SU(3)_f$ symmetry,
the effective Hamiltonian in Eq.~(\ref{Heff}) is transformed as
\begin{eqnarray}\label{Heff2}
{\cal H}_{eff}&=&\frac{G_F}{\sqrt 2}\left[c_- { \epsilon^{ijl} \over 2}H(6)_{lk}+c_+H(\overline{15})_k^{ij}\right]\,,
\end{eqnarray}
where the individual non-zero entries of 
$H(6)_{lk}$ and $H(\overline{15})_k^{ij}$ 
that include $O_\mp$, $O_\mp^\dagger$ and $O'_\mp$ can be presented as 
the matrix forms:
\begin{eqnarray}
H(6)&=&\left(\begin{array}{ccc}
0& 0 & 0\\
0 & 2 & -2s_c\\
0 & -2s_c& 2s_c^2
\end{array}\right)\,,\nonumber\\
H(\overline{15})&=&
\left(\begin{array}{ccc}
\left(\begin{array}{ccc}
0&0&0\\
0&0&0\\
0&0&0
\end{array}\right),
\left(\begin{array}{ccc}
0&s_c&1\\
s_c&0&0\\
1&0&0
\end{array}\right),
\left(\begin{array}{ccc}
0&-s_c^2&-s_c\\
-s_c^2&0&0\\
-s_c&0&0
\end{array}\right)
\end{array}\right).
\end{eqnarray}
Correspondingly, the ${\bf B}_c$ anti-triplet and ${\bf B}_n$ octet states 
are written as
\begin{eqnarray}\label{b_octet}
{\bf B}_{c}&=&(\Xi_c^0,-\Xi_c^+,\Lambda_c^+)\,,\nonumber\\
{\bf B}_n&=&\left(\begin{array}{ccc}
\frac{1}{\sqrt{6}}\Lambda+\frac{1}{\sqrt{2}}\Sigma^0 & \Sigma^+ & p\\
 \Sigma^- &\frac{1}{\sqrt{6}}\Lambda -\frac{1}{\sqrt{2}}\Sigma^0  & n\\
 \Xi^- & \Xi^0 &-\sqrt{\frac{2}{3}}\Lambda
\end{array}\right)\,.
\end{eqnarray}
The adding of the singlet $\eta_1$ to the octet $(\pi,K,\eta_8)$
leads to the nonet of the pseudoscalar meson, given by~\cite{Geng:2017esc}
\begin{eqnarray}\label{M_octet}
M=\left(\begin{array}{ccc}
\frac{1}{\sqrt{2}}(\pi^0+ c_\phi\eta +s_\phi\eta' ) & \pi^- & K^-\\
 \pi^+ & \frac{-1}{\sqrt{2}}(\pi^0- c_\phi\eta -s_\phi\eta') & \bar K^0\\
 K^+ & K^0& -s_\phi\eta +c_\phi\eta'
\end{array}\right)\,,
\end{eqnarray}
where $(\eta,\eta')$ are the mixtures of $(\eta_1,\eta_8)$,
with the mixing angle $\phi=(39.3\pm1.0)^\circ$~\cite{FKS}
for $(c_\phi,s_\phi)=(\cos\phi,\sin\phi)$.

The amplitudes of the ${\bf B}_c\to {\bf B}_n M$ decays
via the effective Hamiltonian in Eq.~(\ref{Heff}) appear to be
${\cal A}({\bf B}_c\to {\bf B}_n M)=\langle {\bf B}_n M|{\cal H}_{eff}|{\bf B}_c\rangle$.
Since ${\cal H}_{eff}$, ${\bf B}_{c(n)}$ and $M$ 
have been in the $SU(3)_f$ forms,
the amplitudes of ${\bf B}_c\to {\bf B}_n M$ 
can be further derived as
\begin{eqnarray}
{\cal A}({\bf B}_c\to {\bf B}_n M)
=\langle {\bf B}_n M|{\cal H}_{eff}|{\bf B}_c\rangle
=\frac{G_F}{\sqrt 2}T({\bf B}_{c}\to {\bf B}_nM)\,,
\end{eqnarray}
with $T({\bf B}_c\to {\bf B}_n M)$ given by~\cite{Lu:2016ogy}
\begin{eqnarray}
 \label{Tamp}
T({\bf B}_c\to {\bf B}_n M)&=&T({\cal O}_6)+T({\cal O}_{\overline{15}})
 \nonumber\\
T({\cal O}_6)&=&
a_1 H_{ij}(6)T^{ik}({\bf B}_n)_k^l (M)_l^j+
a_2 H_{ij}(6)T^{ik}(M)_k^l ({\bf B}_n)_l^j\nonumber\\
&+&
a_3 H_{ij}(6)({\bf B}_n)_k^i (M)_l^j T^{kl}+
h H_{ij}(6)T^{ik}({\bf B}_n)_k^j (M)^l_l\,,\nonumber\\
T({\cal O}_{\overline{15}})&=&
a_4H^{k}_{li}(\overline{15})({\bf B}_{c})^j (M)^i_j ({\bf B}_n)^l_k
+a_5({\bf B}_n)^i_j (M)^l_i H(\overline{15})^{jk}_l ({\bf B}_{c})_k\nonumber\\
&+&
a_6({\bf B}_n)^k_l (M)^i_j H(\overline{15})^{jl}_i ({\bf B}_{c})_k
+a_7({\bf B}_n)^l_i (M)^i_j H(\overline{15})^{jk}_l ({\bf B}_{c})_k\nonumber\\
&+&
h' H_{i}^{jk}(\overline{15})({\bf B}_n)_k^i (M)^l_l ({\bf B}_{c})_j\,,
\end{eqnarray}
where $T_{ij} \equiv ({\bf B}_c)_k\epsilon^{ijk}$, and
$(c_-,c_+)$ have been  absorbed into the $SU(3)$ parameters of
$(a_1,a_2,a_3,h)$ and
$(a_4,a_5,a_6,a_7,h')$, respectively,
and the $h^{(\prime)}$ terms correspond to the contributions
from the singlet $\eta_1$.
%
%
\begin{table}[b!]
\caption{The $T$-amps of the ${\bf B}_{c}\to {\bf B}_n M$ decays.}\label{tab_Tamp}
{
\scriptsize
\begin{tabular}{|c|l|}
\hline
$\Xi_c^0$&\;\;\;\;\;\;\;\;$T$-amp
\\
\hline
$\Sigma^{+} K^{-} $
& $ 2(a_{2}+\frac{a_{4} + a_{7}}{2})$
\\
$\Sigma^{0}\bar{K}^{0}$
&$-\sqrt{2}(a_{2}+a_{3}$
$-\frac{a_{6}-a_{7}}{2})$
\\
$\Xi^{0} \pi^{0} $
& $ -\sqrt{2}(a_{1}-a_{3}$
$-\frac{a_{4}-a_{5}}{2})$
\\
$ \Xi^{0} \eta $ & $ \sqrt{2}c\phi(a_1- a_3 +2h $
$+ \frac{a_4+a_5+2 h'}{2} )$\\
&$- 2s\phi(a_2 +h + \frac{a_7 + h'}{2} ) $ \\
$ \Xi^{0} \eta' $ &  $ \sqrt{2}s\phi(a_1- a_3 +2h $
$+ \frac{a_4+a_5+2 h'}{2} )$\\
&$+ c\phi(a_2 +h + \frac{a_7 + h'}{2} ) $ \\
$\Xi^{-} \pi^{+} $
& $ 2(a_{1}+\frac{a_{5} + a_{6}}{2})$
\\
$\Lambda^{0} \bar{K}^{0} $
&
$-\sqrt{\frac{2}{3}}(2a_1-a_2-a_3$
$+\frac{2a_5-a_6-a_7}{2})$
\\
\hline
$\Sigma^{+} \pi^{-} $
& $-2(a_{2}+\frac{a_{4} + a_{7}}{2})s_c$
\\
$\Sigma^{-} \pi^{+} $
& $-2(a_{1}+\frac{a_{5} + a_{6}}{2})s_c$
\\
$\Sigma^{0} \pi^{0} $
&$-(a_{2}+a_{3}$
$-\frac{a_{4}-a_{5}+a_{6}-a_{7}}{2})s_c$
\\
$ \Sigma^{0} \eta $
& $[-c\phi(a_1+a_2+2h+\frac{a_4+a_5-a_6+a_7+2h'}{2})$\\
&$-\sqrt{2}s\phi(a_3-h-\frac{a_6+h'}{2})]s_c$
\\
$ \Sigma^{0} \eta' $
& $[-s\phi(a_1+a_2+2h+\frac{a_4+a_5-a_6+a_7+2h'}{2})$\\
&$+\sqrt{2}c\phi(a_3-h-\frac{a_6+h'}{2})]s_c$
\\
$\Xi^{-} K^{+} $
& $ 2(a_{1}+\frac{a_{5} + a_{6}}{2})s_c$
\\
$p K^-$
&$2(a_{2}+\frac{a_{4} + a_{7}}{2})s_c$
\\
$\Xi^{0} K^{0} $
& $2(a_{1}-a_2-a_{3}$
$+\frac{a_{5}-a_{7}}{2})s_c$
\\
$n \bar K^{0} $
& $-2(a_{1}-a_{2}-a_{3}+\frac{a_{5}-a_{7}}{2})s_c$
\\
$\Lambda^{0} \pi^{0} $
&
$\sqrt{\frac{1}{3}}(a_1+a_2-2a_3$
\\
&$+\frac{a_4-a_5-a_6-a_7}{2})s_c$
\\
$ \Lambda^{0} \eta $
& $ [\frac{\sqrt{3}c\phi}{3}(a_1+a_2-2a_3+6h $\\
&$+\frac{3a_4+a_5+a_6+a_7+6h'}{2}) $ \\
&$-\frac{\sqrt{6}s\phi}{2}(2a_1+2a_2-a_3+3h $\\
&$+\frac{2a_5-a_6+2a_7+3h'}{2})]s_c$
\\
$ \Lambda^{0} \eta' $
&$[\frac{\sqrt{3}s\phi}{3}(a_1+a_2-2a_3+6h $\\
&$+\frac{3a_4+a_5+a_6+a_7+6h'}{2}) $ \\
&$+\frac{\sqrt{6}c\phi}{2}(2a_1+2a_2-a_3+3h $\\
&$+\frac{2a_5-a_6+2a_7+3h'}{2})]s_c $
\\
\hline
$p\pi^-$&
$-2(a_{2}+\frac{a_{4} + a_{7}}{2})s_c^2$
\\
$\Sigma^{-} K^{+} $
& $-2(a_{1}+\frac{a_{5} + a_{6}}{2})s_c^2$
\\
$\Sigma^{0}{K}^{0}$
&$ \sqrt 2(a_{1}+\frac{a_{5} - a_{6}}{2})s_c^2$
\\
$n \pi^{0} $
&$\sqrt 2(a_{2}-\frac{a_{4} - a_{7}}{2})s_c^2$
\\
$ n \eta $
&$[-\sqrt{2}c\phi( a_2-2h + \frac{a_4 -a_7 -2h'}{2} )$\\
&$+2s\phi( a_1 -a_3 +h +\frac{a_5+h'}{2})]s_c^2$ \\
$ n \eta' $
& $[-\sqrt{2}s\phi( a_2-2h + \frac{a_4 -a_7 -2h'}{2} )$\\
&$-2c\phi( a_1 -a_3 +h +\frac{a_5+h'}{2})]s_c^2$ \\
$\Lambda^{0} {K}^{0} $
&
$-\sqrt{\frac{2}{3}}(a_1-2a_2-2a_3$
$+\frac{a_5+a_6-2a_7}{2})s_c^2$
\\
\hline
\end{tabular}
\begin{tabular}{|c|l|}
\hline
$\Xi_c^+$&\;\;\;\;\;\;\;\;$T$-amp
\\\hline
$\Sigma^{+} \bar{K}^{0} $
&$ -2(a_{3}-\frac{a_{4} + a_{6}}{2})$
\\
$\Xi^{0} \pi^{+} $
& $2(a_{3}+\frac{a_{4} + a_{6}}{2})$
\\[30.5mm]
\hline
$\Sigma^{0} \pi^{+} $
&$\sqrt 2(a_{1}-a_{2}$
\\
&$+\frac{a_{4}-a_{5}+a_{6}+a_{7}}{2})s_c$
\\
$\Sigma^{+} \pi^{0} $
&$-\sqrt 2(a_{1}-a_{2}$
\\
&$-\frac{a_{4}+a_{5}+a_{6}-a_{7}}{2})s_c$
\\
$ \Sigma^{+} \eta $
& $[ \sqrt{2}c\phi(a_1+a_2+2h$ \\
&$-\frac{a_4+a_5+a_6+a_7-2h'}{2})$ \\
&$+2s\phi(a_3-h-\frac{a_6-h'}{2})]s_c$
\\
$ \Sigma^{+} \eta' $
& $[ \sqrt{2}s\phi(a_1+a_2+2h$ \\
&$-\frac{a_4+a_5+a_6+a_7-2h'}{2})$ \\
&$-2c\phi(a_3-h-\frac{a_6-h'}{2})]s_c$
\\
$\Xi^{0} K^{+} $
& $2(a_2+a_{3}+\frac{a_{6} - a_{7}}{2})s_c$
\\
$p \bar K^{0} $
& $2(a_1-a_{3}+\frac{a_{4} - a_{5}}{2})s_c$
\\
$\Lambda^0\pi^+$
& $\sqrt{\frac{2}{3}}(a_1+a_2-2a_3$
\\
&$-\frac{3a_4+a_5+a_6+a_7}{2})s_c$
\\[10mm]
\hline
$\Sigma^{0} {K}^{+} $
&$ \sqrt 2(a_{1}-\frac{a_{5} - a_{6}}{2})s_c^2$
\\
$\Sigma^{+} {K}^{0} $
&$ 2(a_{1}-\frac{a_{5} + a_{6}}{2})s_c^2$
\\
$p \pi^0 $
&$\sqrt 2(a_{2}+\frac{a_{4} - a_{7}}{2})s_c^2$
\\
$ p \eta $
& $[\sqrt{2}c\phi(-a_2 + 2h $ \\
&$+\frac{a_4+a_7 + 2h'}{2})  $ \\
&$ +2s\phi(a_1 - a_3 +h$  \\
&$ - \frac{a_5 - 2h +h'}{2})]s_c$
\\
$ p \eta' $
& $[\sqrt{2}s\phi(-a_2 + 2h $ \\
&$+\frac{a_4+a_7 + 2h'}{2})  $ \\
&$ -2c\phi(a_1 - a_3 +h$  \\
&$ - \frac{a_5 - 2h +h'}{2})]s_c$
\\
$n \pi^{+} $
& $2(a_{2}-\frac{a_{4} + a_{7}}{2})s_c^2$
\\
$\Lambda^0 K^+$
& $\sqrt{\frac{2}{3}}(a_1-2a_2-2a_3$
\\
&$-\frac{a_5+a_6-2a_7}{2})s_c^2$
\\[5mm]
\hline
\end{tabular}
\begin{tabular}{|c|l|}
\hline
$\Lambda_c^+$&\;\;\;\;\;\;\;\; $T$-amp
\\\hline
$\Sigma^{0} \pi^{+} $
&$-\sqrt{2}(a_1-a_2-a_3$
$-\frac{a_5-a_7}{2})$
\\
$\Sigma^{+} \pi^{0} $
& $\sqrt{2}(a_{1}-a_{2}-a_{3}$
$-\frac{a_{5}-a_{7}}{2})$
\\
$ \Sigma^{+} \eta $ & $ \sqrt{2}c\phi(-a_1-a_2+a_3-2h$\\
&$+\frac{a_5+a_7+2h'}{2}) $ \\
&$ +s\phi(-a_4+2h-h')  $ \\
$ \Sigma^{+} \eta' $ & $ \frac{\sqrt{2}s\phi}{2}(-a_1-a_2+a_3-2h$\\
&$+\frac{a_5+a_7+2h'}{2}) $ \\
&$ -c\phi(-a_4+2h-h')  $
\\
$\Xi^{0} K^{+} $
& $-2(a_{2}-\frac{a_{4} + a_{7}}{2})$
\\
$p \bar{K}^{0} $
& $ -2(a_{1}-\frac{a_{5} + a_{6}}{2})$
\\
$\Lambda^{0} \pi^{+} $
&
$-\sqrt{\frac{2}{3}}(a_1+a_2+a_3$
\\
&$-\frac{a_5-2a_6+a_7}{2})$
\\
\hline
$\Sigma^{+} K^{0} $
& $-2(a_{1}-a_{3}-\frac{a_{4}-a_{5}}{2})s_c$
\\
$\Sigma^{0} K^{+} $
&$-\sqrt{2}(a_1-a_3-\frac{a_4+a_5}{2})s_c$
\\
$p \pi^{0} $
& $ -\sqrt 2(a_{2}+a_3-\frac{a_{6} - a_{7}}{2})s_c$
\\
$ p \eta $
&$[\sqrt{2}c\phi(a_2-a_3+2h$\\
&$+\frac{a_6-a_7-2h'}{2})$ \\
&$ +2s\phi(-a_1-h$\\
&$+\frac{a_4+a_5+a_6+h'}{2})]s_c $
\\
$ p \eta' $
& $[\sqrt{2}s\phi(a_2-a_3+2h$\\
&$+\frac{a_6-a_7-2h'}{2})$ \\
&$ -2c\psi(-a_1-h$\\
&$+\frac{a_4+a_5+a_6+h'}{2})]s_c$
\\
$n\pi^+$
&$-2(a_{2}+a_3-\frac{a_{4} + a_{7}}{2})s_c$
\\
$\Lambda^{0} K^{+} $
&
$-\sqrt{\frac{2}{3}}(a_1-2a_2+a_3$
\\
&$-\frac{3a_4-a_5+2a_6+2a_7}{2})s_c$\\[15mm]
\hline
$p {K}^{0} $
& $ 2(a_{3}-\frac{a_{4} + a_{6}}{2})s_c^2$
\\
$nK^+$&
$- 2(a_{3}+\frac{a_{4} + a_{6}}{2})s_c^2$
\\[40.5mm]
\hline
\end{tabular}
}
\end{table}
 With the $T$-amps expanded in Table~\ref{tab_Tamp},
we are enabled to relate
all possible two-body ${\bf B}_c\to {\bf B}_nM$ decays
with the $SU(3)_f$ parameters.
To compute the 
branching ratios, we use the equation 
given by~\cite{pdg}
\begin{eqnarray}\label{p_space}
{\cal B}({\bf B}_c\to {\bf B}_n M)=
\frac{|\vec{p}_{cm}|\tau_{\bf{B}_c}}{8\pi m_{{\bf B}_c}^2 }|{\cal A}({\bf B}_c\to {\bf B}_n M)|^2\,,
\end{eqnarray}
where $|\vec{p}_{cm}|=
\sqrt{[(m_{{\bf B}_c}^2-(m_{{\bf B}_n}+m_M)^2]
[(m_{{\bf B}_c}^2-(m_{{\bf B}_n}-m_M)^2]}/(2 m_{{\bf B}_c})$ and 
$\tau_{\bf{B}_c}$
is the lifetime (the inverse of the total decay width) of  $\bf{B}_c$.
 In Eq.~(\ref{p_space}), the amplitude squared is defined by
\begin{eqnarray}
|{\cal A}({\bf B}_c\to {\bf B}_n M)|^2=\frac{(G_F V_{ij}V_{kl})^2}{2}
T^\dagger({\bf B}_c\to {\bf B}_n M)T({\bf B}_c\to {\bf B}_n M)\,.
\end{eqnarray}
Note that, since the Lorentz indices have been neglected 
in the language of the $SU(3)_f$ symmetry, 
no contractions of the baryon spins are needed, leading to
$T^\dagger({\bf B}_c\to {\bf B}_n M)=T^*({\bf B}_c\to {\bf B}_n M)$.

\section{Numerical Results and Discussions }
For the numerical analysis,
we  note that the contributions of 
the $SU(3)$ parameters
$(a_4,a_5,a_6,a_7,h')$ from $H(\overline{15})$ would be neglected based on the following reasons.
First, 
the contributions to the decay branching rates from $H(\overline{15})$ and $H(6)$
lead to a small ratio of ${\cal R}(\overline{15}/6)=c_+^2/c_-^2\simeq 17\%$
 in terms of  $(c_+,c_-)=(0.76,1.78)$ 
from the QCD calculation at the scale $\mu=1$ GeV
in the naive
dimensional regularization (NDR) scheme~\cite{Li:2012cfa,Fajfer:2002gp}.
Second, 
it is pointed out in Ref.~\cite{Cheng:2018hwl} that $O_+^{(\dagger,\prime)}$
belong to $H(\overline{15})$ in the group structure and 
behave as symmetric operators in color indices, whereas
the baryon wave functions are totally antisymmetric, such that 
the mismatch causes the disappearance of $c_+O_+^{(\dagger,\prime)}$ 
in the calculation of the non-facotrizable effects, which are regarded 
to be significant in the charmed baryon decays.
Note that even though the single ignoring of $H(\overline{15})$ is viable,
a possible interference between the amplitudes with
$H(6)$ and $H(\overline{15})$ may be sizable to fail this assumption, 
which will be tested in the fit.
Hence, being from $H(6)$ 
the parameters ($a_1,a_2,a_3,h$) in Eq.~(\ref{Tamp})  are kept for the fit,
which are in fact complex.
Since an overall phase can be removed without losing generality,
we set $a_1$ to be real,  such that
there are seven real independent parameters to be determined,
given by
\begin{eqnarray}\label{7p}
a_1, a_2e^{i\delta_{a_2}},a_3e^{i\delta_{a_3}},he^{i\delta_h}\,.
\end{eqnarray}
We use the minimum $\chi^2$ fit for the determination,
given by
\begin{eqnarray}
\chi^2=
\sum_{i} \bigg(\frac{{\cal B}^i_{th}-{\cal B}^i_{ex}}{\sigma_{ex}^i}\bigg)^2+
\sum_{j}\bigg(\frac{{\cal R}^j_{th}-{\cal R}^j_{ex}}{\sigma_{ex}^j}\bigg)^2\,,
\end{eqnarray}
where ${\cal B}_{th}^i$  and ${\cal R}_{th}^j$ 
stand for the separated  decay branching ratios and
the ratios of the two-decay branching fractions from the $SU(3)$ amplitudes, 
while ${\cal B}_{ex}^i$ and ${\cal R}_{ex}^j$ are the corresponding 
experimental data, 
along with $\sigma^{i}_{ex}$ and $\sigma^{j}_{ex}$ the $1\sigma$ uncertainties,
respectively.
With the ten experimental data in Table~\ref{data},
the global fit results in
\begin{eqnarray}\label{su3_fit}
&&(a_1,a_2,a_3,h)=(0.244\pm 0.006,0.115\pm 0.014,0.088\pm 0.019,0.105\pm0.073)\,\text{GeV}^3\,,\nonumber\\
&&(\delta_{a_2},\delta_{a_3},\delta_h)=(78.1\pm 7.1, 35.1\pm 8.7,10.2\pm29.6)^\circ\,,\nonumber\\
&&\chi^2/d.o.f=5.32/3=1.77\,,
\end{eqnarray}
where $d.o.f$ represents the degree of freedom.
The numerical values for the parameters in Eq.~(\ref{su3_fit}) are
 the theoretical inputs, which are used to predict
the two-body ${\bf B}_c\to {\bf B}$ decays in Table~\ref{tab_result}.

\begin{table}[b!]
\caption{The data of the ${\bf B}_c\to {\bf B}_n M$ decays.}\label{data}
\begin{tabular}{|c||c|}
\hline
Branching ratios
&Data~\cite{pdg,
Ablikim:2017ors} \\
\hline
$10^2{\cal B}(\Lambda_c^+ \to p \bar K^0)$
&$3.16\pm 0.16$\\
$10^2{\cal B}(\Lambda_c^+ \to \Lambda \pi^+)$
&$1.30\pm 0.07$\\
$10^2{\cal B}(\Lambda_c^+ \to \Sigma^+ \pi^0)$
& $1.24\pm 0.10$\\
$10^2{\cal B}(\Lambda_c^+ \to \Sigma^0 \pi^+)$
& $1.29\pm 0.07$\\
$10^2{\cal B}(\Lambda_c^+ \to \Xi^0 K^+)$
& $0.50\pm 0.12$\\
\hline
\end{tabular}
\begin{tabular}{|c||c|}
\hline
Branching ratios
&Data~\cite{pdg,
Ablikim:2017ors} \\
\hline
$10^2{\cal B}(\Lambda_c^+ \to \Sigma^+ \eta)$
&$0.70\pm 0.23$ \\
$10^4{\cal B}(\Lambda_c^+ \to \Lambda K^+)$
&$6.1\pm 1.2$\\
$10^4{\cal B}(\Lambda_c^+ \to \Sigma^0 K^+)$
&$5.2\pm 0.8$ \\
$10^4{\cal B}(\Lambda_c^+ \to p \eta)$
&$12.4\pm 3.0$ \\
${\cal R}=\frac{{\cal B}( \Xi^0_c \to \Lambda \bar K^0)}{{\cal B}(\Xi^0_c \to \Xi^-\pi^+)}$
&$0.420\pm 0.056$\\
\hline
\end{tabular}
\end{table}
%
%
\begin{table}[b!]
\caption{The numerical results of
the ${\bf B}_{c}\to {\bf B}_n M$ decays
with ${\cal B}_{{\bf B}_nM}\equiv {\cal B}({\bf B}_c\to {\bf B}_nM)$,
where the number  with the dagger ($\dagger$)
is the reproduction of the experimental data input,
instead of the prediction.}\label{tab_result}
{
\scriptsize
\begin{tabular}{|c|c|c|}
\hline
$\Xi_c^0$&our results&Ref.~\cite{Wang:2017gxe}
\\
\hline
$10^3{\cal B}_{\Sigma^{+} K^{-}}$
&$3.5 \pm 0.9$&$3.1 \pm 0.9$\\
$10^3{\cal B}_{\Sigma^{0} \bar{K}^{0}}$
&$4.7 \pm 1.2$&$4.6\pm 1.4$\\
$10^3{\cal B}_{\Xi^{0} \pi^{0}}$
&$4.3 \pm 0.9$&$0.7-18.1$ \\
$10^3{\cal B}_{\Xi^{0} \eta}$
&$1.7 ^{+ 1.0 }_{- 1.7 }$& \\
$10^3{\cal B}_{\Xi^{0} \eta'}$
&$8.6 ^{+ 11.0 }_{-\;\,6.3 }$& \\
$10^3{\cal B}_{\Xi^{-} \pi^{+}}$
&$15.7 \pm 0.7$&$22.4 \pm 3.4$ \\
$10^3{\cal B}_{\Lambda^{0} \bar{K}^{0}}$
&$8.3 \pm 0.9$&$9.4\pm 1.6$ \\
\hline
$10^4{\cal B}_{\Sigma^{+} \pi^{-}}$
&$2.0 \pm 0.5$& \\
$10^4{\cal B}_{\Sigma^{-} \pi^{+}}$
&$9.0 \pm 0.4$& \\
$10^4{\cal B}_{\Sigma^{0} \pi^{0}}$
&$3.2 \pm 0.3$& \\
$10^4{\cal B}_{\Sigma^{0} \eta}$
&$3.6 ^{+ 1.0 }_{- 0.9 }$& \\
$10^4{\cal B}_{\Sigma^{0} \eta'}$
&$1.7 ^{+ 3.0 }_{- 1.5 }$& \\
$10^4{\cal B}_{\Xi^{-} K^{+}}$
&$7.6 \pm 0.4$& \\
$10^4{\cal B}_{\Xi^{0} K^{0}}$
&$6.3 \pm 1.2$& \\
$10^4{\cal B}_{p K^{-}}$
&$2.1 \pm 0.5$& \\
$10^4{\cal B}_{n \bar{K}^{0}}$
&$7.9 \pm 1.4$& \\
$10^4{\cal B}_{\Lambda^{0} \pi^{0}}$
&$0.2 \pm 0.2$& \\
$10^4{\cal B}_{\Lambda^{0} \eta}$
&$1.6 ^{+ 1.2 }_{- 0.8 }$& \\
$10^4{\cal B}_{\Lambda^{0} \eta'}$
&$9.4 ^{+ 11.6 }_{-\;\,6.8 }$& \\
\hline
$10^6{\cal B}_{p \pi^{-}}$
&$12.1 \pm 3.1$& \\
$10^6{\cal B}_{\Sigma^{-} K^{+}}$
&$44.5 \pm 2.1$& \\
$10^6{\cal B}_{\Sigma^{0} K^{0}}$
&$22.3 \pm 1.0$& \\
$10^6{\cal B}_{n \pi^{0}}$
&$6.0 \pm 1.5$& \\
$10^6{\cal B}_{n \eta}$
&$26.5^{+ 11.4 }_{- 10.1}$& \\
$10^6{\cal B}_{n \eta'}$
&$ 30.7 ^{+ 42.3 }_{- 24.4 }$& \\
$10^6{\cal B}_{\Lambda^{0} K^{0}}$
&$14.4 \pm 3.7$& \\
\hline
\end{tabular}
\begin{tabular}{|c|c|c|}
\hline
$\Xi_c^+$&our results&Ref.~\cite{Wang:2017gxe}
\\
\hline
$10^3{\cal B}_{\Sigma^{+} \bar{K}^{0}}$
&$8.0 \pm 3.9$&$0.1-102.2$ \\
$10^3{\cal B}_{\Xi^{0} \pi^{+}}$
&$8.1 \pm 4.0 $&$1.2-96.8$ \\[25mm]
\hline
$10^4{\cal B}_{\Sigma^{0} \pi^{+}}$
&$18.5 \pm 2.2$& \\
$10^4{\cal B}_{\Sigma^{+} \pi^{0}}$
&$18.5 \pm 2.2$& \\
$10^4{\cal B}_{\Sigma^{+} \eta}$
&$28.4 ^{+ 8.2 }_{- 6.9 }$& \\
$10^4{\cal B}_{\Sigma^{+} \eta'}$
&$13.2 ^{+ 24.0 }_{- 11.9 }$& \\
$10^4{\cal B}_{\Xi^{0} K^{+}}$
&$18.0 \pm 4.7$& \\
$10^4{\cal B}_{p \bar{K}^{0}}$
&$20.3 \pm 4.2$& \\
$10^4{\cal B}_{\Lambda^{0} \pi^{+}}$
&$1.6 \pm 1.2$& \\[25mm]
\hline
$10^5{\cal B}_{\Sigma^{0} K^{+}}$
&$8.8 \pm 0.4$& \\
$10^5{\cal B}_{\Sigma^{+} K^{0}}$
&$17.6 \pm 0.8$& \\
$10^6{\cal B}_{p \pi^{0}}$
&$23.8 \pm 6.1$& \\
$10^5{\cal B}_{p \eta}$
&$10.5 ^{+ 4.5 }_{- 4.0 }$& \\
$10^5{\cal B}_{p \eta'}$
&$12.1 ^{+ 16.7 }_{-\;\,9.7 }$& \\
$10^6{\cal B}_{n \pi^{+}}$
&$47.6 \pm 12.2$& \\
$10^6{\cal B}_{\Lambda^{0} K^{+}}$
&$56.8 \pm 14.5$& \\
\hline
\end{tabular}
\begin{tabular}{|c|c|c|}
\hline
$\Lambda_c^+$&our results&Ref.~\cite{Wang:2017gxe}
\\
\hline
$10^3{\cal B}_{\Sigma^{0} \pi^{+}}$
&$(1.3 \pm 0.2)^\dagger$&$(1.27 \pm 0.17)^\dagger$ \\
$10^3{\cal B}_{\Sigma^{+} \pi^{0}}$
&$(1.3 \pm 0.2)^\dagger$&$(1.27 \pm 0.17)^\dagger$  \\
$10^2{\cal B}_{\Sigma^{+} \eta}$
&$(0.7 ^{+0.4}_{-0.3})^\dagger$& \\
$10^2{\cal B}_{\Sigma^{+} \eta'}$
&$1.0 ^{+1.6}_{-0.8}$& \\
$10^2{\cal B}_{\Xi^{0} K^{+}}$
&$(0.5 \pm 0.1)^\dagger$&$(0.50 \pm 0.12)^\dagger$ \\
$10^2{\cal B}_{p \bar{K}^{0}}$
&$(3.3 \pm 0.2)^\dagger$&$(2.72-3.60)^\dagger$ \\
$10^2{\cal B}_{\Lambda^{0} \pi^{+}}$
&$(1.3\pm 0.2)^\dagger$&$(1.30\pm 0.17)^\dagger$ \\
\hline
$10^4{\cal B}_{\Sigma^{+} K^{0}}$
&$8.0 \pm 1.6$& \\
$10^4{\cal B}_{\Sigma^{0} K^{+}}$
&$(4.0 \pm 0.8)^\dagger$& \\
$10^4{\cal B}_{p \pi^{0}}$
&$5.7 \pm 1.5$& \\
$10^4{\cal B}_{p \eta}$
&$(12.5 ^{+ 3.8 }_{- 3.6 })^\dagger$& \\
$10^4{\cal B}_{p \eta'}$
&$12.2 ^{+ 14.3 }_{-\;\,8.7 }$& \\
$10^4{\cal B}_{n \pi^{+}}$
&$11.3 \pm 2.9$& \\
$10^4{\cal B}_{\Lambda^{0} K^{+}}$
&$(4.6 \pm 0.9)^\dagger$& \\[25mm]
\hline
$10^6{\cal B}_{p K^{0}}$
&$12.2 \pm 6.0$& \\
$10^6{\cal B}_{n K^{+}}$
&$12.2 \pm 6.0$& \\[25mm]
\hline
\end{tabular}
}
\end{table}
%

Since the value of $\chi^2/d.o.f\simeq 1.8$ in Eq.~(\ref{su3_fit}) indicates a good fit,
 there exists no inconstancy by neglecting  $H(\overline{15})$ in our analysis.
Note that the determinations of $|a_1|$ and $|a_2|$ depend on
$T(\Lambda_c^+\to p\bar K^0)=-2a_{1}$
and
$T(\Lambda_c^+\to \Xi^{0} K^{+})=-2a_{2}$
in Table~\ref{tab_Tamp}, respectively, 
by ignoring $(a_{5} + a_{6})$ and $(a_{4} + a_{7})$, associated with $H(\overline{15})$.
Similarly, one can extract $|a_3|$ based on $T(\Xi_c^+\to \Xi^{0} \pi^{+} )=2a_{3}+(a_{4} + a_{6})\simeq 2a_3$.
Consequently, we get
\begin{eqnarray}\label{ignoring}
&R_0{\cal B}(\Lambda_c^+\to p\bar K^0)
={\cal B}(\Xi_c^0\to \Xi^-\pi^+)
&=(15.7 \pm 0.7)\times 10^{-3}\,,\nonumber\\
&R_0{\cal B}(\Lambda_c^+\to \Xi^{0} K^{+})
={\cal B}(\Xi_c^0\to\Sigma^{+} K^{-})
&=(0.4 \pm 0.1)\times 10^{-2}\,,\nonumber\\
&{\cal B}(\Xi_c^+\to\Sigma^{+} \bar{K}^{0})
={\cal B}(\Xi_c^+\to\Xi^{0} \pi^{+})
&=(8.1 \pm 4.0)\times 10^{-3}\,,
\end{eqnarray}
without the contributions from $H(\overline{15})$,
where $R_0=\tau_{\Xi_c^0}/\tau_{\Lambda_c^+}=0.56\pm 0.07$.
To check if the $H(\overline{15})$ terms are indeed negligible, we may use
the relations from Table~\ref{tab_Tamp}, given by
\begin{eqnarray}\label{re_Tamp}
&T(\Lambda_c^+\to p\bar K^0)+T(\Xi_c^0\to \Xi^-\pi^+)
&=2(a_{5} + a_{6})\,,\nonumber\\
&T(\Lambda_c^+\to \Xi^{0} K^{+})+T(\Xi_c^0\to\Sigma^{+} K^{-})
&=2(a_{4} + a_{7})\,,\nonumber\\
&T(\Xi_c^+\to\Xi^{0} \pi^{+})+T(\Xi_c^+\to\Sigma^{+} \bar{K}^{0})
&=2(a_{4} + a_{6})\,.
\end{eqnarray}
Clearly, if the results in Eq.~(\ref{ignoring})
do not  agree with the future measurements, 
the contributions from $H(\overline{15})$ should be reconsidered as seen in Eq.~(\ref{re_Tamp}).

According to the PDG~\cite{pdg},
the branching fractions in the $\Xi_c^0$ decays
are observed to be relative to
${\cal B}_{\Xi^-\pi^+}\equiv {\cal B}(\Xi_c^0\to \Xi^-\pi^+)$,
 predicted in Table~\ref{tab_result}.
Hence, by using the partial observations of
${\cal B}(\Xi_c^0\to\Lambda K^-\pi^+)=(1.07\pm 0.14){\cal B}_{\Xi^-\pi^+}$,
${\cal B}(\Xi_c^0\to\Lambda K^+K^-)=(0.029\pm 0.007){\cal B}_{\Xi^-\pi^+}$, and
${\cal B}(\Xi_c^0\to\Xi^- e^+\nu_e)=(3.1\pm 1.1){\cal B}_{\Xi^-\pi^+}$, we obtain
\begin{eqnarray}
&{\cal B}(\Xi_c^0\to\Lambda K^-\pi^+)&=(16.8\pm 2.3)\times 10^{-3}\,,\nonumber\\
&{\cal B}(\Xi_c^0\to\Lambda K^+K^-)&=(4.5\pm 1.1)\times 10^{-4}\,,\nonumber\\
&{\cal B}(\Xi_c^0\to\Xi^- e^+\nu_e)&=(48.7\pm 17.4)\times 10^{-3}\,.
\end{eqnarray}
Similarly,
the branching fractions in the $\Xi_c^+$ decays
are measured to be relative to ${\cal B}(\Xi_c^+\to \Xi^-\pi^+\pi^+)$,
which has not been theoretically and experimentally studied yet.
With ${\cal B}(\Xi_c^+\to \Xi^0\pi^+)/{\cal B}(\Xi_c^+\to\Xi^-\pi^+\pi^+)
=0.55\pm 0.16$~\cite{pdg} and ${\cal B}(\Xi_c^+\to \Xi^0\pi^+)$ in Table~\ref{tab_result},
we find
\begin{eqnarray}
{\cal B}_{\Xi^-2\pi^+}\equiv {\cal B}(\Xi_c^+\to\Xi^-\pi^+\pi^+)=(14.7\pm 8.4 )\times 10^{-3}\,.
\end{eqnarray}
Subsequently, the relative branching fractions of
${\cal B}(\Xi_c^+\to p K_s^0 K_s^0)=(0.087\pm 0.021){\cal B}_{\Xi^-2\pi^+}$,
${\cal B}(\Xi_c^+\to \Sigma^+ K^-\pi^+)=(0.94\pm 0.10){\cal B}_{\Xi^-2\pi^+}$,
${\cal B}(\Xi_c^+\to \Xi^0\pi^+\pi^0)=(2.3\pm 0.7){\cal B}_{\Xi^-2\pi^+}$ and
${\cal B}(\Xi_c^+\to\Xi^0 e^+\nu_e)=(2.3^{+0.7}_{-0.8}){\cal B}_{\Xi^-2\pi^+}$~\cite{pdg}
lead to
\begin{eqnarray}
&{\cal B}(\Xi_c^+\to p K_s^0 K_s^0)&=(1.3\pm 0.8)\times 10^{-3}\,,\nonumber\\
&{\cal B}(\Xi_c^+\to \Sigma^+ K^-\pi^+)&=(13.8\pm 8.0)\times 10^{-3}\,,\nonumber\\
&{\cal B}(\Xi_c^+\to \Xi^0\pi^+\pi^0)&=(33.8\pm 21.9)\times 10^{-3}\,,\nonumber\\
&{\cal B}(\Xi_c^+\to\Xi^0 e^+\nu_e)&=(33.8^{+21.9}_{-22.6})\times 10^{-3}\,.
\end{eqnarray}

By adding the $h^{(\prime)}$ terms,
we are able to include the contributions from the singlet $\eta_1$
in the $SU(3)_f$ amplitudes, which have been used to explain
the observations of
${\cal B}(\Lambda_c^+ \to \Sigma^+ \eta)$ and
${\cal B}(\Lambda_c^+ \to p \eta)$. Nonetheless,
the estimations  of ${\cal B}(\Lambda_c^+ \to \Sigma^+(p) \eta')\simeq
{\cal B}(\Lambda_c^+ \to \Sigma^+(p) \eta)$~\cite{Geng:2017esc} 
 show  no inequality as
${\cal B}(B \to K \eta')\gg{\cal B}(B \to K \eta)$ or
${\cal B}(B \to K^* \eta)\gg{\cal B}(B \to K^* \eta')$.
On the other hand, it is interesting to note that, despite of the large uncertainties,
the $\Xi_c\to{\bf B}_n \eta^{(\prime)}$ decays contain
the similar inequalities between the $\eta$ and $\eta'$ modes, given by
\begin{eqnarray}
&{\cal B}(\Xi_c^0\to \Xi^{0} \eta,\Xi^{0} \eta')
&=(1.7 ^{+ 1.0 }_{- 1.7 },8.6 ^{+ 11.0 }_{-\;\,6.3 })\times 10^{-3}\,,\nonumber\\
&{\cal B}(\Xi_c^0\to \Lambda^{0} \eta,\Lambda^{0} \eta')
&=(1.6 ^{+ 1.2 }_{- 0.8 },9.4 ^{+ 11.6 }_{-\;\,6.8 })\times 10^{-4}\,,\nonumber\\
&{\cal B}(\Xi_c^+\to \Sigma^{+} \eta,\Sigma^{+} \eta')
&=(28.4 ^{+ 8.2 }_{- 6.9 },13.2 ^{+ 24.0 }_{- 11.9 })\times 10^{-4}\,.
\end{eqnarray}
We remark that 
as shown  in Table~\ref{tab_result}, 
our numerical results for 
the Cabibbo-allowed processes are consistent with those in Ref.~\cite{Wang:2017gxe},
where 
${\cal B}({\bf B}_c\to {\bf B}_n \bar K^0)$ are taken from 
${\cal B}({\bf B}_c\to {\bf B}_n K^0_S)$. 
Finally, we 
emphasize that there is a  discrepancy  between 
the theory and data for ${\cal B}(\Lambda_c^+\to p\pi^0)$.
In Table~\ref{tab_result}, 
${\cal B}(\Lambda_c^+\to p\pi^0)$ is predicted to be $(5.7 \pm 1.5)\times 10^{-4}$, 
whereas it is
measured to be less than $3\times 10^{-4}$~\cite{Ablikim:2017ors}.
Nonetheless, 
the estimation in the factorization approach also gives
${\cal B}(\Lambda_c^+\to p\pi^0)=f_{\pi}^2/(2f_K^2) s_c^2\,{\cal B}(\Lambda_c^+\to p\bar K^0)
=(5.5\pm 0.3)\times 10^{-4}$ to be as large as our $SU(3)_f$ prediction in Table~\ref{tab_result} ,
with the experimental input of ${\cal B}(\Lambda_c^+\to p\bar K^0)=
(3.16\pm 0.16)\times 10^{-2}$~\cite{pdg}.
Clearly, to resolve this inconsistency,
 it is necessary to re-measure the decay of $\Lambda_c^+\to p\pi^0$ in the future experiment.

\section{Conclusion}
With the $SU(3)_f$ symmetry,
we have studied the two-body anti-triplet charmed baryon weak decays.
We have predicted that
${\cal B}(\Xi_c^0\to \Xi^-\pi^+)=(15.7 \pm 0.7)\times 10^{-3}$ and
${\cal B}(\Xi_c^+\to\Xi^-\pi^+\pi^+)=(14.7\pm 8.4)\times 10^{-3}$,
while the branching ratios of
the $\Xi_c^0$ and $\Xi_c^+$ decays are measured to
be relative to ${\cal B}(\Xi_c^0\to \Xi^-\pi^+)$ and
${\cal B}(\Xi_c^+\to\Xi^-\pi^+\pi^+)$, respectively.
Hence, we have extracted that
${\cal B}(\Xi_c^0\to\Lambda K^-\pi^+,\Lambda K^+K^-,\Xi^- e^+\nu_e)=(16.8\pm 2.3,0.45\pm 0.11,48.7\pm 17.4)\times 10^{-3}$
and
${\cal B}(\Xi_c^+\to p K_s^0 K_s^0,\Sigma^+ K^-\pi^+,\Xi^0\pi^+\pi^0,\Xi^0 e^+\nu_e)=
(1.3\pm 0.8,13.8\pm 8.0,33.8\pm 21.9,33.8^{+21.9}_{-22.6})\times 10^{-3}$.
In addition, we have shown that
${\cal B}(\Xi_c^0\to \Xi^{0} \eta,\Xi^{0} \eta')=
(1.7 ^{+ 1.0 }_{- 1.7 },8.6 ^{+ 11.0 }_{-\;\,6.3 })\times 10^{-3}$,
${\cal B}(\Xi_c^0\to \Lambda^{0} \eta,\Lambda^{0} \eta')
=(1.6 ^{+ 1.2 }_{- 0.8 },9.4 ^{+ 11.6 }_{-\;\,6.8 })\times 10^{-4}$ and
${\cal B}(\Xi_c^+\to \Sigma^{+} \eta,\Sigma^{+} \eta')
=(28.4 ^{+ 8.2 }_{- 6.9 },13.2 ^{+ 24.0 }_{- 11.9 })\times 10^{-4}$,
representing the inequalities, similar to those of
${\cal B}(B \to K \eta')\gg{\cal B}(B \to K \eta)$ or
${\cal B}(B \to K^* \eta)\gg{\cal B}(B \to K^* \eta')$ in the mesonic $B$ decays involving $\eta^{(')}$.
According to our predictions, the branching ratios of 
two and three-body $\Xi_c$ decays 
are accessible to the experiments at BESIII and LHCb.

\section*{ACKNOWLEDGMENTS}
This work was supported in part by National Center for Theoretical Sciences,
MoST (MoST-104-2112-M-007-003-MY3), and
National Science Foundation of China (11675030).

\end{document}